\documentclass[seceq]{ptptex}

\usepackage{graphicx}
\usepackage{wrapft}
\usepackage{bm}

\newcommand{\ket}[1]{|#1\rangle}

\notypesetlogo                       

\title{
Separation of a Slater determinant wave function with a neck structure into spatially localized subsystems
}

\author{
Yasutaka \textsc{Taniguchi}$^1$ and Yoshiko \textsc{Kanada-En'yo}$^2$%
}

\inst{
$^1$RIKEN Nishina Center for Accelerator-Based Science, RIKEN, Wako, Saitama 351-0109, Japan.\\
$^2$Department of Physics, Kyoto University, Kyoto, Kyoto 606-8502, Japan.
}

\abst{
A method to separate a Slater determinant wave function with a two-center neck structure into spatially localized subsystems is
proposed, and its potential applications are presented.
An orthonormal set of spatially localized single-particle wave functions is obtained by diagonalizing the coordinate operator for the major axis of a necked system.
Using the localized single-particle wave functions, the wave function of each subsystem is defined.
Therefore, defined subsystem wave functions are used to obtain density distributions, mass centers, and energies of subsystems.
The present method is applied to separations of Margenau--Brink cluster wave functions of 
$\alpha + \alpha$, $^{16}$O + $^{16}$O, and $\alpha + ^{16}$O into their subsystems, and also to separations of antisymmetrized molecular dynamics wave functions of $^{10}$Be into $\alpha$ + $^6$He subsystems.
The method is simple and applicable to the separation of general Slater determinant wave functions that have neck structures into subsystem wave functions.
}

\begin{document}

\maketitle

\section{Introduction}

Nuclear systems often exhibit multi-center structures having spatially localized subsystems,
as seen in phenomena such as cluster structures and fusion/fission processes.
In cluster structures, each system consists of more than one spatially localized subsystems, called clusters, 
and intercluster motion is a degree of freedom used to describe the structure of a total system.
In such multi-center systems, constituent nucleons of each cluster and the internuclear distance are useful parameters to specify cluster features. 
Identification of subsystems is also important for nuclear reactions,
in which colliding nuclei are subsystems.
In order to have an in-depth understanding of nuclear systems with subsystems, wave functions of each subsystem should be defined microscopically.

For microscopic wave functions of quantum fermion systems,  
the Slater determinant description is often adopted as used 
in the Hartree--Fock (HF), the HF--Bogoliubov (HFB), and the time-dependent HF approaches, the Brink--Bloch model, and the antisymmetrized molecular dynamics (AMD) method\cite{amd_ono1,amd_ono2,amd_enyo}. 
Many applications to studies of nuclear structures and reactions show that
two-center structures have localized subsystems. 
However, a method to microscopically separate a Slater determinant into subsystem wave functions is not obvious,
particularly when subsystems overlap with each other.
One of the popular ways to separate the density of a total system into those of subsystems is to employ the method of sharp-cut density separated by a boundary plane.
However, the sharp-cut density is based on a classical picture and it is a nonphysical entity in quantum systems 
because density distributions of subsystems have singularities at the boundary plane.
Recently, a method to localize quasi-particle states by unitary transformation applied to single-particle orbits so as to maximize difference of occupations of left and right of the neck position in the HFB method (YG method)\cite{younes}.
In the YG method, localized single-particle wave functions depend on the neck position, which has ambiguities.
In the cases of Margenau--Brink (MB) wave functions\cite{mb}, which are used in the Brink--Bloch model, and AMD wave functions, subsystem wave functions can be broadly defined as cluster wave functions or 
groups of single-particle Gaussian wave packets. However, the subsystem wave functions 
in the overlapping region of two clusters contain 
nonphysical components that vanish because of the Pauli blocking effect, that is, antisymmetrization of nucleons
in the total wave function.
Saraceno \textit{et al.} proposed canonical variables of the distance and relative momentum of the mass centers of the two clusters in an $\alpha$--$\alpha$ system including the antisymmetrization effect between the $\alpha$ clusters (S method).
However, wave functions of subsystems are not defined in the method\cite{saraceno}, and it is not applicable to systems that have one or more open-shell clusters.

The aim of this paper is to propose a method applicable for any systems to define wave functions of spatially localized subsystems in a Slater determinant wave function by a linear transformation of single-particle orbits using an operator with respect to the internuclear relative coordinate.
With this method, spatially localized single-particle wave functions are defined with no assumption and a total system is microscopically separated into subsystems.
Applications to $N = Z$ nuclei, $^8$Be ($\alpha + \alpha$), $^{32}$S ($^{16}$O + $^{16}$O), $^{20}$Ne ($\alpha + ^{16}$O), and $N \neq Z$ nucleus, $^{10}$Be ($\alpha + ^6$He) are shown.
We compare the internuclear distances defined by the present method with those defined by sharp-cut density, single-particle Gaussian centroids of AMD wave functions, and the S method.
Energies of subsystems are also discussed.

In \S\ref{framework}, we propose a method to define wave functions of subsystems.
In \S\ref{results}, the method is applied to MB wave functions for $N = Z$ systems
and AMD wave functions of $^{10}$Be,
and the density distributions of subsystems, internuclear distance, and energies of subsystems are shown.
In \S\ref{discussions}, characteristics of the method are discussed.
Finally, conclusions are given in \S\ref{conclusions}.

\section{Framework}
\label{framework}

Suppose a Slater determinant wave function $| \mathrm{\Phi} \rangle$ having a two-center structure as
\begin{equation}
 | \mathrm{\Phi} \rangle = \hat{\cal A} | \tilde{\varphi}_1, \tilde{\varphi}_2, ......, \tilde{\varphi}_A \rangle,\label{local1}
\end{equation}
where $\hat{\cal A}$ is the antisymmetrization operator, $A$ is the mass number of a total system, and $\tilde{\varphi}_1$, ..., $\title{\varphi}_A$ are single-particle wave functions.
Here, the $z$ axis is chosen to be the major axis of the total system.
Note that the total wave function $| \mathrm{\Phi} \rangle$ is invariant except for the normalization under any linear transformations of single-particle wave functions 
$\{\tilde{\varphi}_i\}\rightarrow \{{\varphi}_i\}$ that give nonzero determinants.

In order to separate the total system into two subsystems I and II, subsystem 
wave functions are defined as follows:
First, single-particle wave functions are 
transformed to diagonalize norm and coordinate operators $\hat{z}$,
\begin{eqnarray}
 & | \varphi_i \rangle = \sum_j | \tilde{\varphi}_j \rangle c_{ji},& \\
 & \langle \varphi_i | \varphi_j \rangle = \delta_{ij},&\\
 & \langle \varphi_i|\hat{z}|\varphi_j\rangle = z_i\delta_{ij},&\label{local4}
\end{eqnarray}
where $z_1\le z_2\le \cdots \le z_A$. 
The eigenvalue $z_i$ indicates the mean position of the $i$-th nucleon. 
If a single-particle wave function $\varphi_i$ is localized at a
certain position, $z_i$ can be regarded as the nucleon position in a semi-classical picture.
Therefore, it is natural to classify $\varphi_i$ into two groups 
$\{\varphi_1, \varphi_2, \cdots, \varphi_{A_1} \}$ and
$\{\varphi_{A_1 + 1}, \varphi_{A_1 + 2}, \cdots, \varphi_A \}$
according to the distribution of $z_i$.
Then, wave functions $|\mathrm{\Phi}_\mathrm{I} \rangle$ and $|\mathrm{\Phi}_\mathrm{II} \rangle$ of the subsystems I and II, respectively, are described as
\begin{equation}
 |\mathrm{\Phi}_\mathrm{I}  \rangle = \hat{\cal A} |\varphi_1, \varphi_2, \cdots, \varphi_{A_1}\rangle,
\end{equation}
\begin{equation}
 |\mathrm{\Phi}_\mathrm{II} \rangle = \hat{\cal A} |\varphi_{A_1 + 1}, \varphi_{A_1 + 2}, \cdots, \varphi_A\rangle.
\end{equation}
Here, $A_1$ and $A_2\equiv A-A_1$ are regarded as the number of nucleons constituting subsystems I and II, respectively.

Once single-particle wave functions are separated, the internuclear distance and expectation values of any operators (such as the density distribution and energy for each subsystem) can be calculated.
The mass centers $\mathbf{R}_\mathrm{I}$ and $\mathbf{R}_\mathrm{II}$ of subsystems I and II are obtained as 
\begin{eqnarray}
 \mathbf{R}_\mathrm{I} &=& \frac{1}{A_1} \sum_{i=1,\cdots,A_1}
  \langle \varphi_i | \hat{\mathbf{r}} | \varphi_i \rangle, \\ 
 \mathbf{R}_\mathrm{II} &=& \frac{1}{A_2} \sum_{i=A_1+1,\cdots,A}
  \langle \varphi_i | \hat{\mathbf{r}} | \varphi_i \rangle,
\end{eqnarray}
respectively.
Then, the internuclear distance $R$ is defined as the distance between the mass centers,
\begin{equation}
 R = | \mathbf{R}_\mathrm{II} - \mathbf{R}_\mathrm{I} |.
\end{equation}
Density distributions $\rho_\mathrm{I}(\mathbf{r})$ and $\rho_\mathrm{II}(\mathbf{r})$ of subsystems I and II are calculated by
\begin{eqnarray}
\rho_\mathrm{I}({\bf r})&=& \sum_{i=1,\cdots,A_1} \rho_i(\mathbf{r}),\\
\rho_\mathrm{II}({\bf r})&=& \sum_{i=A_1+1,\cdots,A} \rho_i(\mathbf{r}),
\end{eqnarray}
respectively.
Here, $\rho_i(\mathbf{r})$ is the density distribution of the $i$-th nucleon,
\begin{equation}
  \rho_i(\mathbf{r}) = \langle \varphi_i|\delta(\hat{\bf r} - {\bf r})\ket{\varphi_i}.
\end{equation}

Hamiltonian $\hat{H}_\mathrm{n}$ for the subsystem $\mathrm{n}$ is defined as,
\begin{equation}
 \hat{H}_\mathrm{n} \equiv \hat{T}_\mathrm{n} + \hat{V}_{\mathrm{n}} - \hat{T}_{\mathrm{Gn}},
\end{equation}
where $\hat{T}_\mathrm{n}$, $\hat{V}_{\mathrm{n}}$, and $\hat{T}_{\mathrm{Gn}}$ are the kinetic, potential, and mass-center motion terms, respectively, for the subsystem $\mathrm{n}$.
The kinetic term $\hat{T}'_\mathrm{n} \equiv \hat{T}_\mathrm{n} - \hat{T}_{\mathrm{Gn}}$ with the center-of-mass correction is described as a two-body operator (see Appendix).
Energy $E_\mathrm{n}$ of the subsystems $\mathrm{n}$ is obtained as
\begin{equation}
 E_\mathrm{n} = \langle \mathrm{\Phi_n} | \hat{H}_\mathrm{n} | \mathrm{\Phi_n} \rangle.
\end{equation}
In the present study, the Volkov No.~2 force is used as the effective interaction, and the Majorana parameter  is set to 0.59 to adjust the ground-state energy of $^{16}$O.

\section{Results}
\label{results}

\subsection{$N = Z$ nuclei}

In this section, the current separation method is applied to the MB wave functions.
In the MB wave function for a system consisting of two clusters, 
clusters I and II are expressed by the shell-model configurations with centers at $(0, 0, - \frac{A_2}{A} d)$ and $(0, 0, + \frac{A_1}{A} d)$,
respectively.
The parameter $d$ specifies the degree of cluster development.
MB wave functions for two-cluster systems, $\alpha + \alpha$, $\alpha + ^{16}\mathrm{O}$, and $^{16}\mathrm{O} + ^{16}\mathrm{O}$, are considered, and subsystem density distributions, internuclear distances, and the energy of each cluster are calculated by the present method.

\subsubsection{$^8$Be ($\alpha + \alpha$)}

In a MB wave function $| \mathrm{\Phi}_{\alpha + \alpha} \rangle$ 
with an $\alpha + \alpha$ cluster structure, 
the $\alpha$ clusters I and II are expressed by the 
$(0s)^4$ shell model wave function shifted to the positions $-d/2$ and $d/2$, respectively, as
\begin{eqnarray}
 \ket{\mathrm{\Phi}_{\alpha + \alpha}}& = &\hat{\cal A} \ket{\tilde\varphi_1,\  \tilde\varphi_2,\cdots,\tilde\varphi_8}, \\
 \ket{\tilde\varphi_{1, 5}}& = &\ket{\tilde\phi_{1,2}\otimes p\uparrow }, \\
 \ket{\tilde\varphi_{2, 6}}& = &\ket{\tilde\phi_{1,2}\otimes p\downarrow }, \\
 \ket{\tilde\varphi_{3, 7}}& = &\ket{\tilde\phi_{1,2}\otimes n\uparrow }, \\
 \ket{\tilde\varphi_{4, 8}}& = &\ket{\tilde\phi_{1,2}\otimes n\downarrow }, \\
 \langle\mathbf{r}|{\tilde\phi_{1,2}}\rangle & = &\left( \frac{2\nu}{\pi} \right)^{\frac{3}{4}} e^{- \nu \left( {\bf r}\pm \frac{\bf d}{2} \right)^2 },
\end{eqnarray}
 where the ${\bf d}=(0,0,d)$, $p$ and $n$ are a proton and neutron, respectively, the $\uparrow$ and $\downarrow$ indicate up and down spins, respectively, and the $\nu$ indicates the width parameter of Gaussian wave packets. 
In the large $d$ limit, the antisymmetrization effect is small enough and the mean position of $\alpha$ clusters I and II are $\mp \mathbf{d} / 2$, respectively.
 Therefore, in a semi-classical picture, the parameter $d$ is regarded as the internuclear distance.
 However, when the spatial parts $\tilde{\phi}_{1}$ and $\tilde{\phi}_{2}$ have a non-negligible overlap with each other, $d$ does not represent the internuclear distance because the single-particle wave functions in the overlapping region are modified by the Pauli blocking effect.

The transformation given by Eq.~(\ref{local1})--(\ref{local4}) obtains analytical forms of spatially localized single-particle wave functions in the $\alpha + \alpha$ system as
\begin{eqnarray}
 \ket{\varphi_{1, 5}}& = &\ket{\phi_{1,2}\otimes p\uparrow }, \\
 \ket{\varphi_{2, 6}}& = &\ket{\phi_{1,2}\otimes p\downarrow }, \\
 \ket{\varphi_{3, 7}}& = &\ket{\phi_{1,2}\otimes n\uparrow }, \\
 \ket{\varphi_{4, 8}}& = &\ket{\phi_{1,2}\otimes n\downarrow }, \\
 \langle\mathbf{r}|{\phi_{1,2}}\rangle
  & = &
  c_+ e^{- \nu \left(\mathbf{r} \pm \frac{\mathbf{d}}{2}\right)^2} + c_-  e^{- \nu \left(\mathbf{r} \mp \frac{\mathbf{d}}{2}\right)^2}, \\
 c_\pm
 &=&
 \frac{1}{2} \left( \frac{2 \nu}{\pi} \right)^\frac{3}{4} 
 \left(
  \sqrt{\frac{1 - e^{- \frac{1}{2} \nu d^2}}{1 - e^{- \nu d^2}}}
  \pm
  \sqrt{\frac{1 + e^{- \frac{1}{2} \nu d^2}}{1 - e^{- \nu d^2}}}
 \right).
\end{eqnarray}
Eigen values $\{z_i\}$ of $\hat{z}$ are
\begin{eqnarray}
 z_{1, 2, 3, 4} &=& - \frac{d}{2 \sqrt{1 - e^{- \nu d^2}}},\\
 z_{5, 6, 7, 8} &=& + \frac{d}{2 \sqrt{1 - e^{- \nu d^2}}},
\end{eqnarray}
which are scaled by $1/{\sqrt{1 - e^{- \nu d^2}}}$ when compared to the positions of centroids of wave packets $\mp \mathbf{d} / 2$.
The internuclear distance $R$ between $\alpha$ clusters is, therefore, 
\begin{equation}
 R = \frac{d}{\sqrt{1 - e^{- \nu d^2}}}.
\end{equation}
The internuclear distance $R$ tends to $\nu^{-1/2}$ in the $d \rightarrow 0$ limit.

\begin{figure}[tbp]
 \begin{center}
  \begin{tabular}{cc}
   \includegraphics[width=0.45\textwidth]{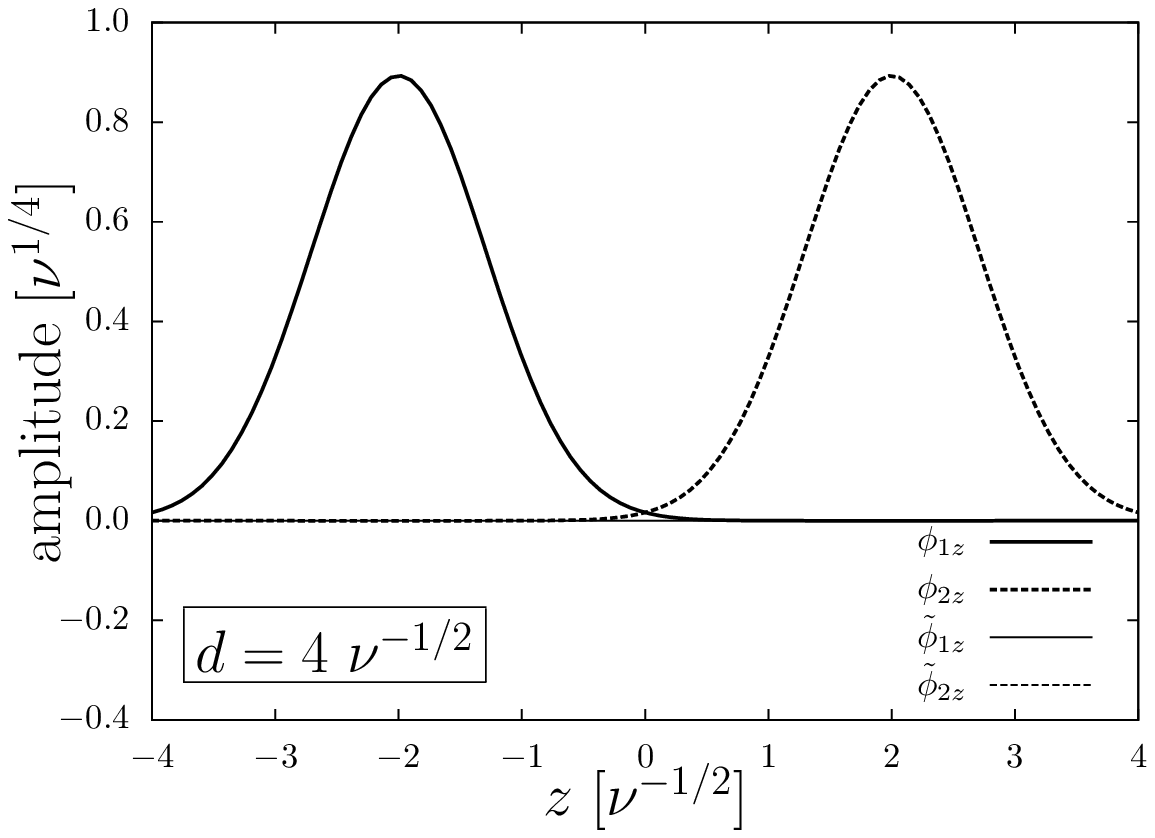}
   &
   \includegraphics[width=0.45\textwidth]{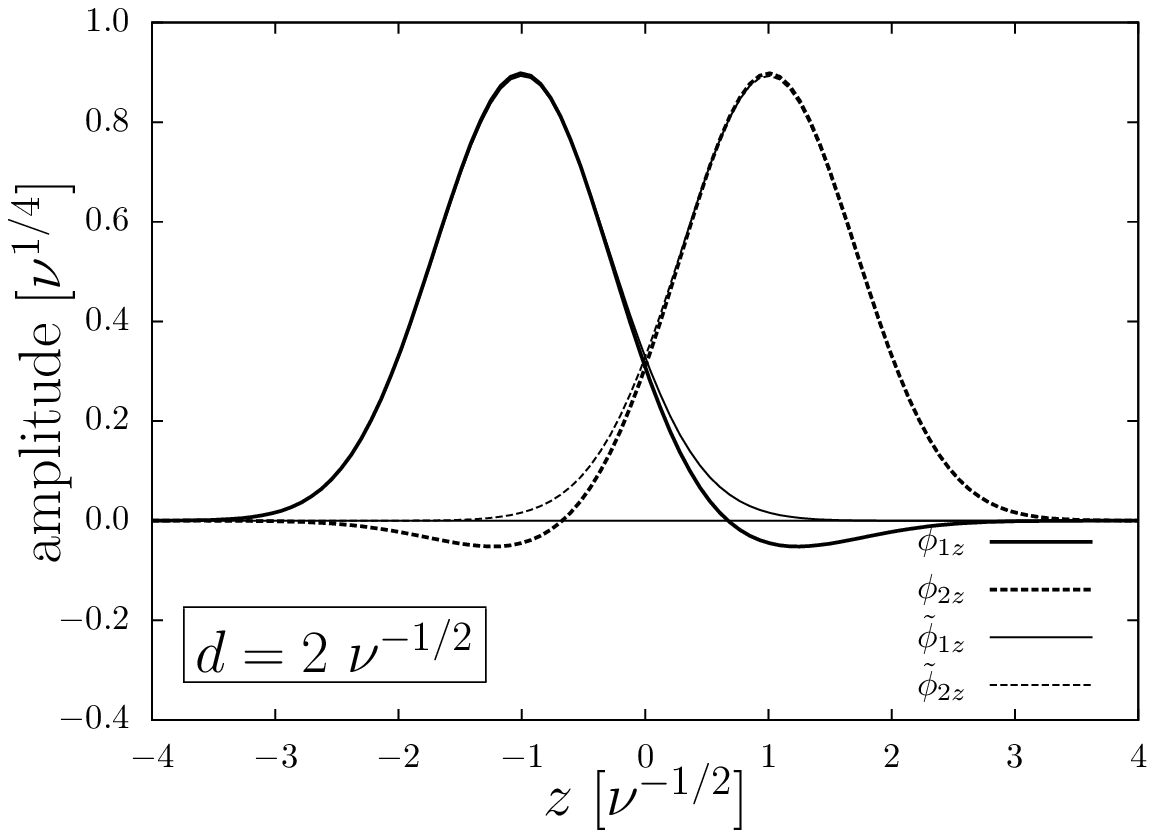}
   \\
   \includegraphics[width=0.45\textwidth]{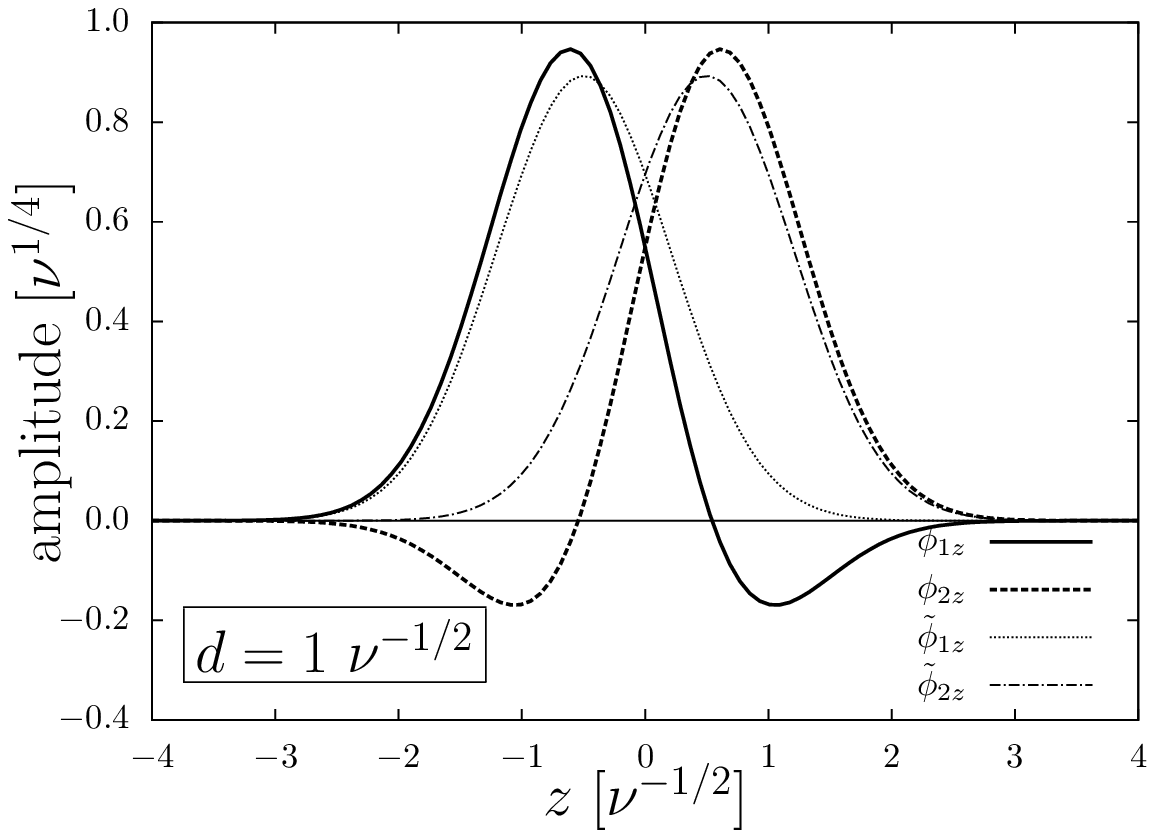}
   &
       \includegraphics[width=0.45\textwidth]{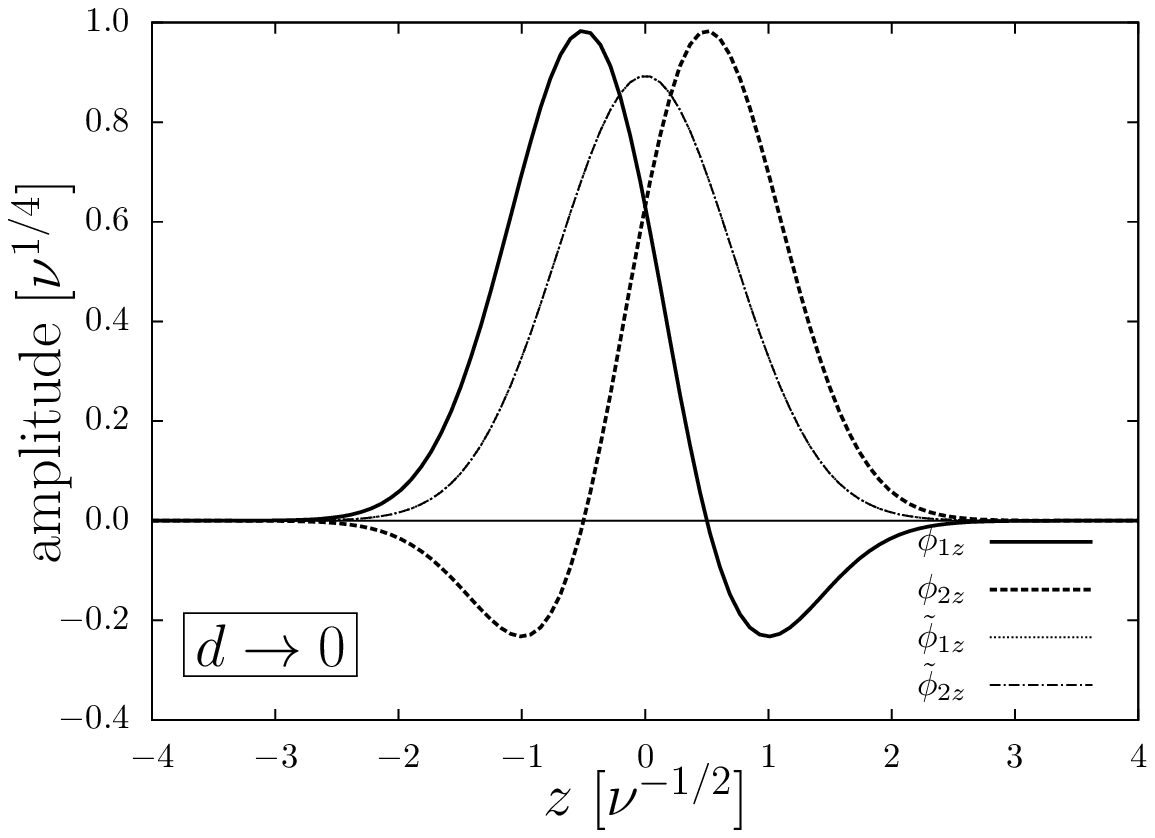}
       \\
  \end{tabular}
  \caption{
  Amplitudes of a $z$-part of single-particle wave functions $\phi_1$ (solid) and $\phi_2$ (dashed), where a $z$-part of $\phi_i$ is defined as $\phi_i / \left[(2 \nu / \pi)^{1/2} e^{-\nu ( x^2 + y^2)}\right]$,  in $\alpha + \alpha$ MB wave functions are plotted  with the parameters $d = 4$ (upper left), 2 (upper right), and $1~\nu^{-1/2}$ (lower left), while $d \rightarrow 0$ (lower right) are plotted as functions of $z$.
  The Gaussian wave packets for the original single particle wave functions $\tilde\phi_{1, 2z}$ (dotted and dot-dashed, respectively) before antisymmetrization are also shown.
  Units of the amplitudes and $z$ are in $\nu^{1/4}$ and $\nu^{-1/2}$, respectively.
  }
  \label{wf_alpha_alpha}
 \end{center}
\end{figure}

Figure~\ref{wf_alpha_alpha} shows the spatial parts of the spatially localized single-particle wave functions $\phi_{1,2}$ as well as Gaussian wave packets $\tilde\phi_{1,2}$ at $d = 4$, 2, and $1~\nu^{-1/2}$, and the $d\rightarrow 0$ limit. 
In the case of $d = 4 ~\nu^{-1/2}$, $\phi_i$ almost coincides with $\tilde\phi_i$,
whereas $\phi_i$ has an obvious node in the $d \leq 2\nu^{-1/2}$ region where 
$\tilde\phi_1$ and $\tilde\phi_2$ overlap.
The wave functions $\phi_1$ and $\phi_2$ are spatially localized even for $d \rightarrow 0$.

    \begin{figure}[tbp]
    \begin{center}
  \begin{tabular}{cc}
   \includegraphics[width=0.45\textwidth]{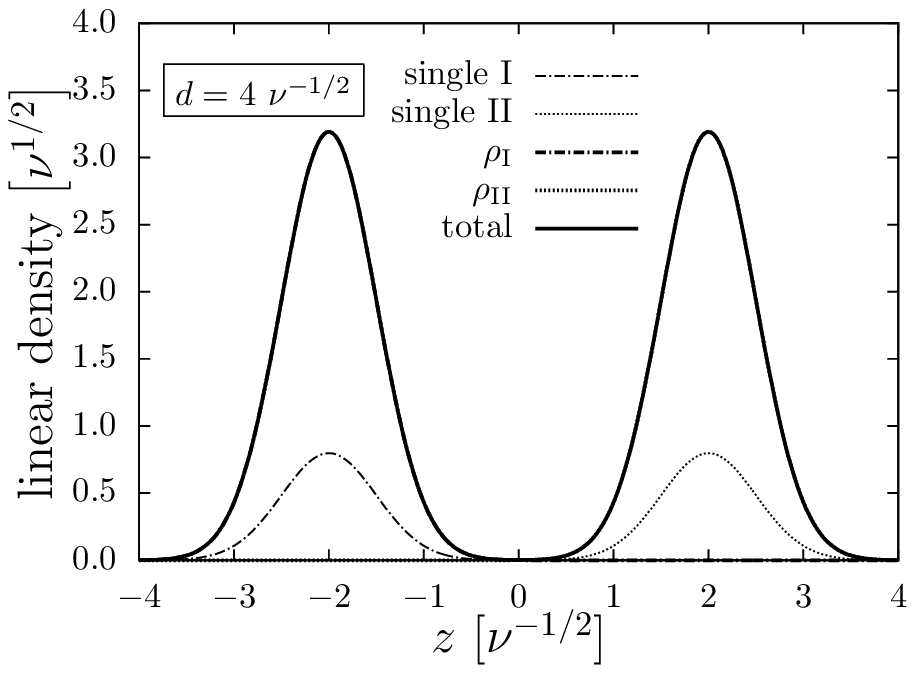}
   &
   \includegraphics[width=0.45\textwidth]{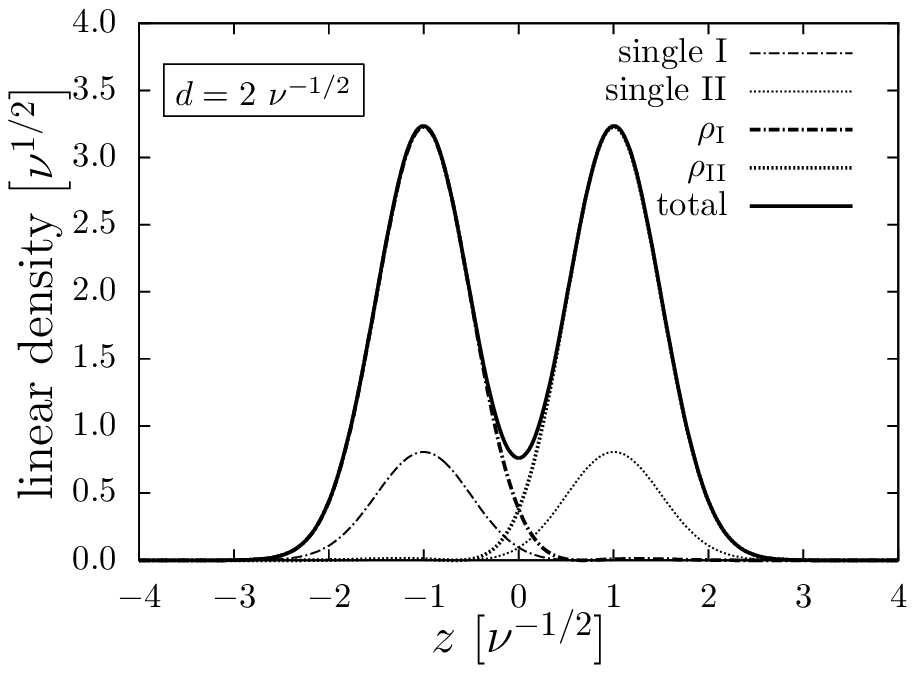}
   \\
   \includegraphics[width=0.45\textwidth]{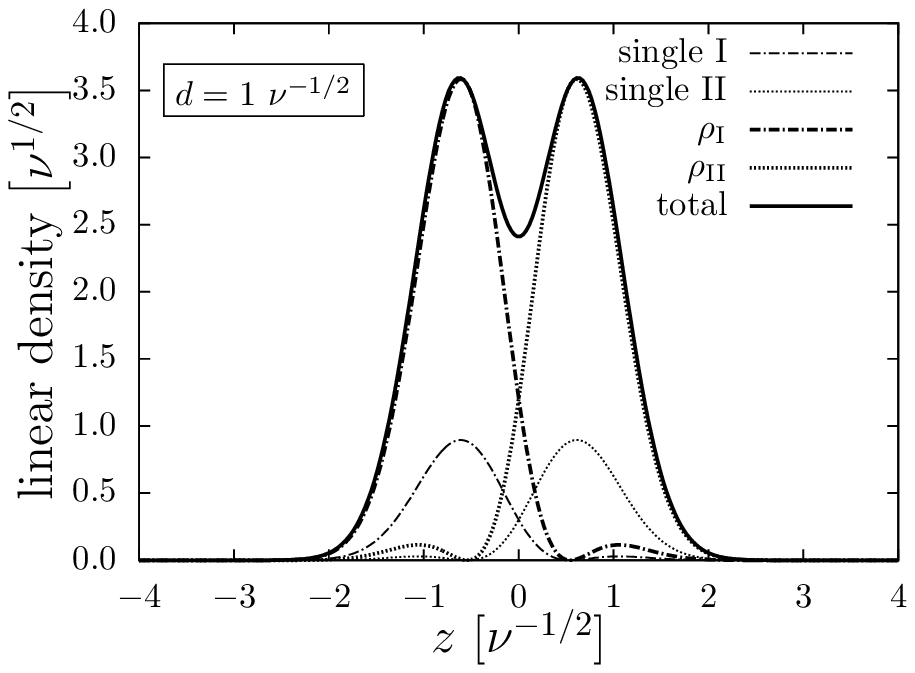}
   &
       \includegraphics[width=0.45\textwidth]{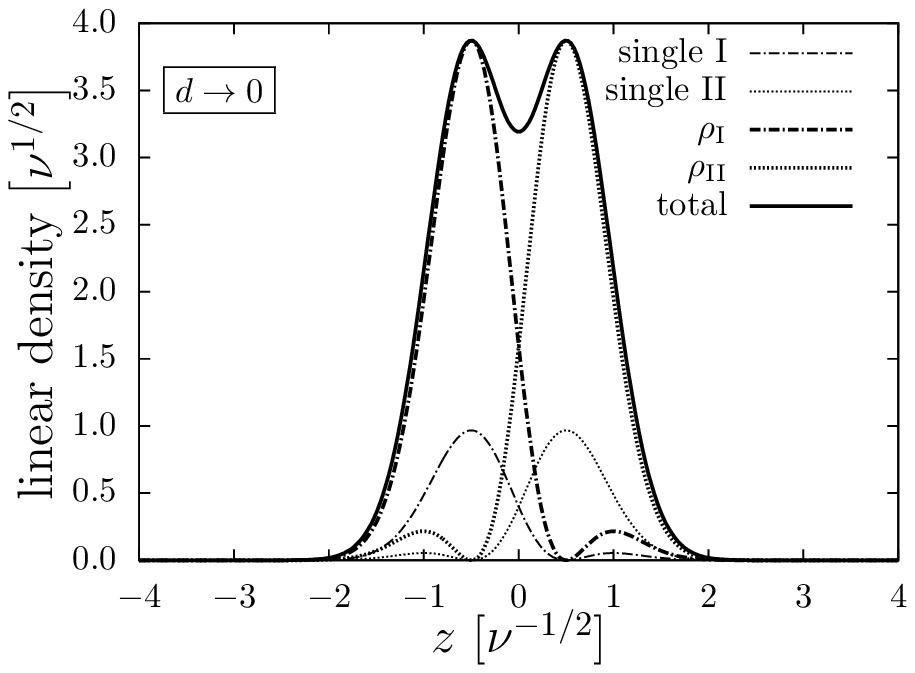}
       \\
  \end{tabular}

     \caption{
     Linear density distributions of $\alpha$ clusters I and II, spatially localized single-particle wave functions, and the total system in $\alpha + \alpha$ MB wave functions with the parameter $d= 4$ (upper left), 2 (upper right), $1~\nu^{-1/2}$ (lower left), and $d \rightarrow 0$ (lower right) are plotted as functions of $z$.
     Thin dot-dashed and dotted lines represent single-particle wave functions for $\alpha$ clusters I and II, respectively.
     Thick dot-dashed and dotted lines represent $\alpha$ clusters I and II, respectively.
     Solid lines represent total wave functions.
     Units of linear densities and $z$ are in $\nu^{1/2}$ and $\nu^{-1/2}$, respectively.
     }
    \label{density_alpha_alpha}
    \end{center}
    \end{figure}

Figure~\ref{density_alpha_alpha} shows linear density distributions $\rho_{xy\mathrm{I, II}} (z)$ of $\alpha$ clusters I and II defined by the integrated subsystem densities with respect to $x$ and $y$,
\begin{equation}
 \rho_{xy\mathrm{I, II}} (z) = \int\!\!\!\!\int\!\! dx dy~\rho_{\mathrm{I, II}} (\mathbf{r}).
\end{equation}
Density distributions of four single-particle wave functions in each cluster are the same because of spin and isospin saturation.
Unlike the sharp-cut separation, subsystem densities are always smooth.
Density distributions of $\alpha$ clusters I and II are spatially localized, even in the case of $d \rightarrow 0$.

\begin{figure}[tbp]
 \begin{center}
  \begin{tabular}{cc}
   \begin{minipage}{0.45\textwidth}
    \includegraphics[width=\textwidth]{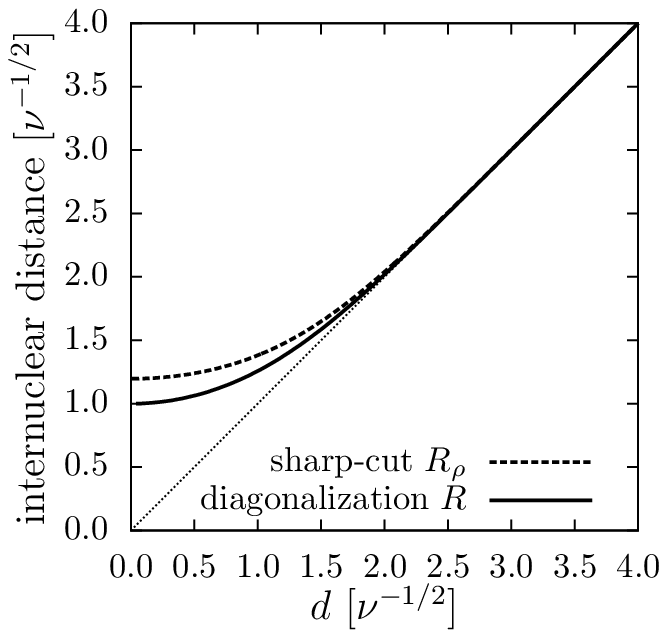}
    \caption{
    Internuclear distance defined by the present method (solid) and sharp-cut separation (dashed) as functions of distance between centroids of wave packets in units of $\nu^{-1/2}$.
    }
    \label{distance_2alpha}
   \end{minipage}&
   \begin{minipage}{0.45\textwidth}
    \includegraphics[width=\textwidth]{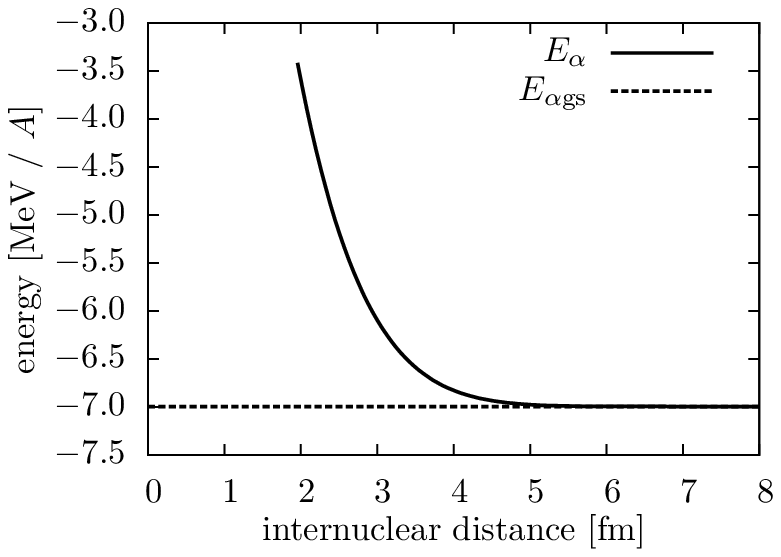}
    \caption{
    Energies of the $\alpha$ cluster in $^8$Be are plotted as a function of internuclear distance in the unit of MeV/$A$ (solid), where $A$ is the mass number of each cluster.
    The dashed line shows the ground-state energy of the $\alpha$ cluster.
    }
    \label{energy_AA}
   \end{minipage}
  \end{tabular}
 \end{center}
\end{figure}

Figure~\ref{distance_2alpha} shows internuclear distances $R$ by the present method and  $R_\rho$ obtained by the sharp-cut separation as functions of distance parameter $d$ 
between the centroids of wave packets.
$R_\rho$ is calculated analytically as
\begin{equation}
 R_\rho = \frac{1}{\sqrt{2 \pi \nu }} \frac{1}{\sinh\left(\frac{1}{2} \nu d^2\right)} \left[ \sum_{n = 0}^\infty \frac{n! \left(2 \nu d^2\right)^n}{(2n)!} - e^{-\frac{1}{2}\nu d^2} \right].
\end{equation}
In $d \gtrsim 2 ~\nu^{-1/2}$, $R$ and $R_\rho$ are similar and they are close to $d$.
In $d \lesssim 2 ~\nu^{-1/2}$, where $\alpha$ clusters overlap with each other, 
$R$ is smaller than $R_\rho$. 

Figure~\ref{energy_AA} shows energies of the $\alpha$ clusters in the $^8$Be as function of internuclear distance $R$.
Width parameter $\nu$ is set to 0.27 fm$^{-2}$.
In $R \gtrsim 5$ fm region, the $\alpha$ clusters are in the ground state, but in $R \lesssim 5$ fm region, the $\alpha$ clusters are distorted by the antisymmetrization effect between clusters, and their excitation energies go up smoothly.

\subsubsection{$^{20}$Ne ($\alpha + ^{16}$O)}

\begin{figure}[tbp]
 \begin{center}
  \begin{tabular}{cc}
   \includegraphics[width=0.45\textwidth]{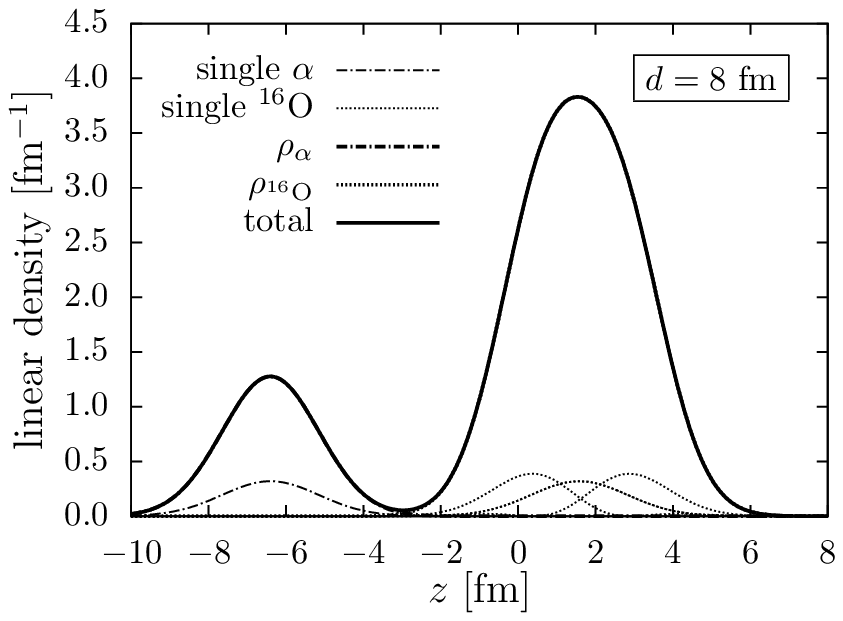} &
   \includegraphics[width=0.45\textwidth]{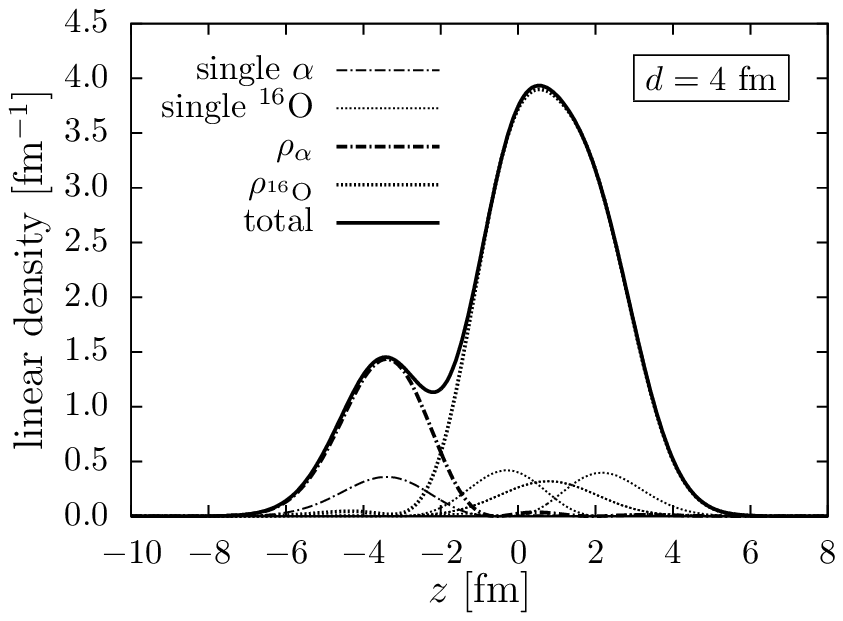} \\
   \includegraphics[width=0.45\textwidth]{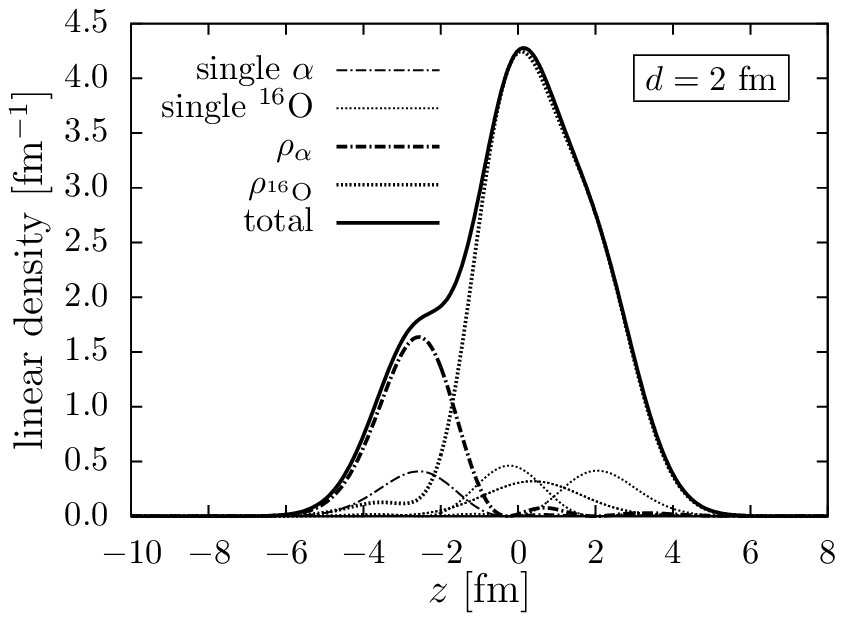} &
   \includegraphics[width=0.45\textwidth]{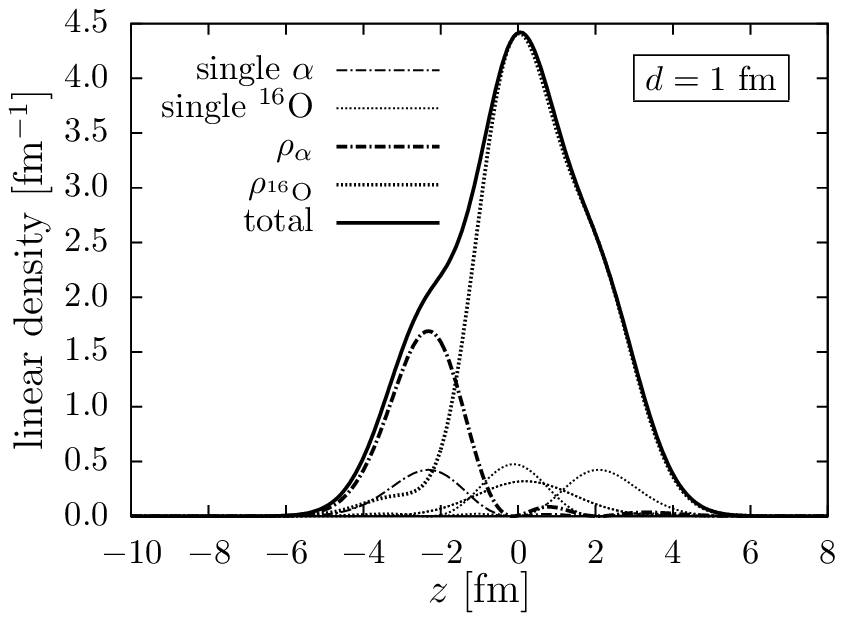}
  \end{tabular}
  \caption{
  Linear density distributions of $\alpha$ and $^{16}$O clusters and those for the total system in the $\alpha + ^{16}$O MB  wave functions
 with the parameter $d= 8$, 4, 2, and 1 fm.
  Thin dot-dashed and dotted lines represent single-particle wave functions for $\alpha$ and $^{16}$O clusters, respectively.
  Thick dot-dashed and dotted lines represent $\alpha$ and $^{16}$O clusters, respectively.
  Solid lines represent total wave functions.
  Units of linear densities and $z$ are in fm$^{-1}$ and fm, respectively.
  }
  \label{density_AO}
 \end{center}
\end{figure}

Figure~\ref{density_AO} shows linear density distributions for the $\alpha$ and $^{16}$O clusters, as well as spatially localized single-particle orbits of $\alpha$ and $^{16}$O clusters in $\alpha + ^{16}$O MB wave functions with $d = 8, 4, 2$, and 1 fm.
Width parameter $\nu$ is set to 0.16 fm$^{-2}$.
At the small $d$ region, $\alpha$ and $^{16}$O clusters have nodes in the overlap region because of antisymmetrization effects between clusters, but they are still spatially localized.
Subsystem densities have no singularity.

\begin{figure}[tbp]
 \begin{center}
  \begin{tabular}{cc}
   \begin{minipage}{0.45\textwidth}
    \includegraphics[width=\textwidth]{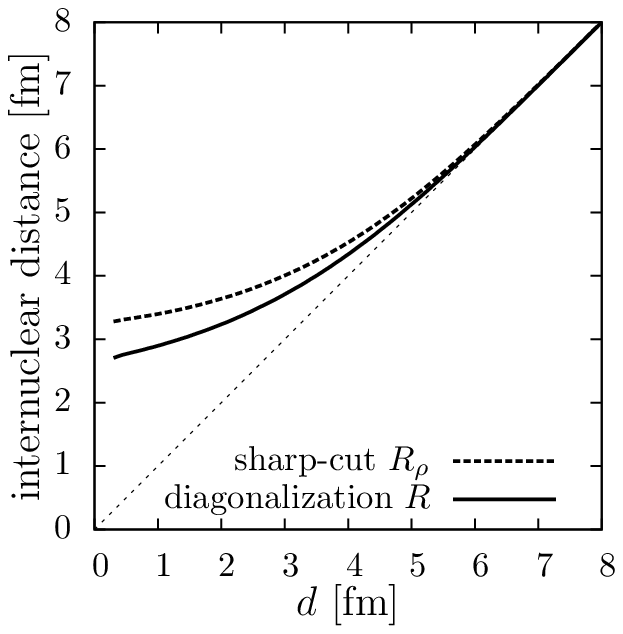}
    \caption{
    Internuclear distance between $\alpha$ and $^{16}$O clusters defined by the present method (solid) and sharp-cut separation (dashed) as functions of distance between centroids of wave packets.
    Units are in fm.
    }
    \label{distance_AO}
   \end{minipage}&
   \begin{minipage}{0.45\textwidth}
    \includegraphics[width=\textwidth]{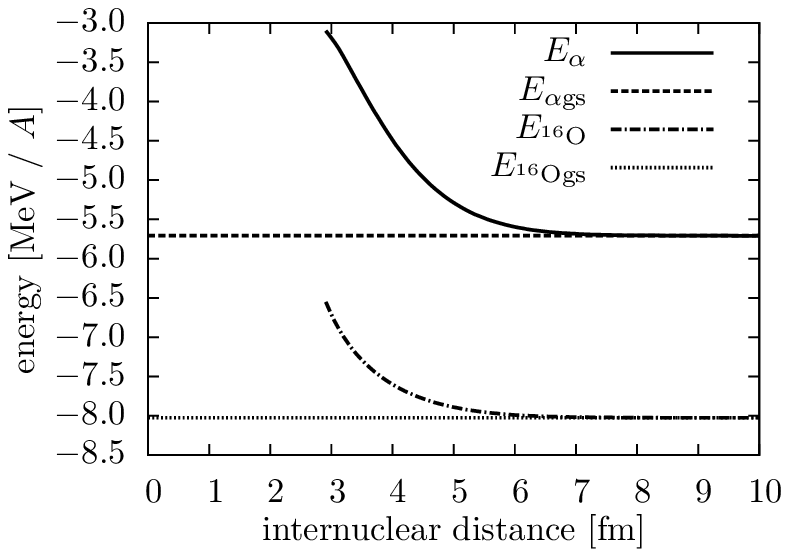}
    \caption{
    Energies of the $\alpha$ (solid) and $^{16}$O (dot-dashed) clusters in $^{20}$Ne are plotted as functions of internuclear distance in the unit of MeV/$A$, where $A$ is the mass number of the cluster.
    The dashed and dotted lines shows the ground-state energy of the $\alpha$ and $^{16}$O clusters, respectively.
    }
    \label{energy_AO}
   \end{minipage}
\end{tabular} 
 \end{center}
\end{figure}

Figure~\ref{distance_AO} shows the internuclear distance $R$ and $R_\rho$ between $\alpha$ and $^{16}$O clusters defined by the present method and sharp-cut separation, respectively.
$R$ is smaller than $R_\rho$ in the $d \lesssim 6$ fm region, and they are similar in the $d \gtrsim 6$ fm region. 

Figure~\ref{energy_AO} shows energies of the $\alpha$ and $^{16}$O clusters in the $^{20}$Ne as functions of internuclear distance $R$.
In $R \gtrsim 6$ fm region, the $\alpha$ and $^{16}$O clusters are in the ground state, but in $R \lesssim 6$ fm region, the $\alpha$ and $^{16}$O clusters are distorted by the antisymmetrization effect between clusters, and their excitation energies go up.

\subsubsection{$^{32}$S ($^{16}\mathrm{O}+^{16}$O)}

    \begin{figure}[tbp]
     \begin{center}
      \begin{tabular}{cc}
       \includegraphics[width=0.45\textwidth]{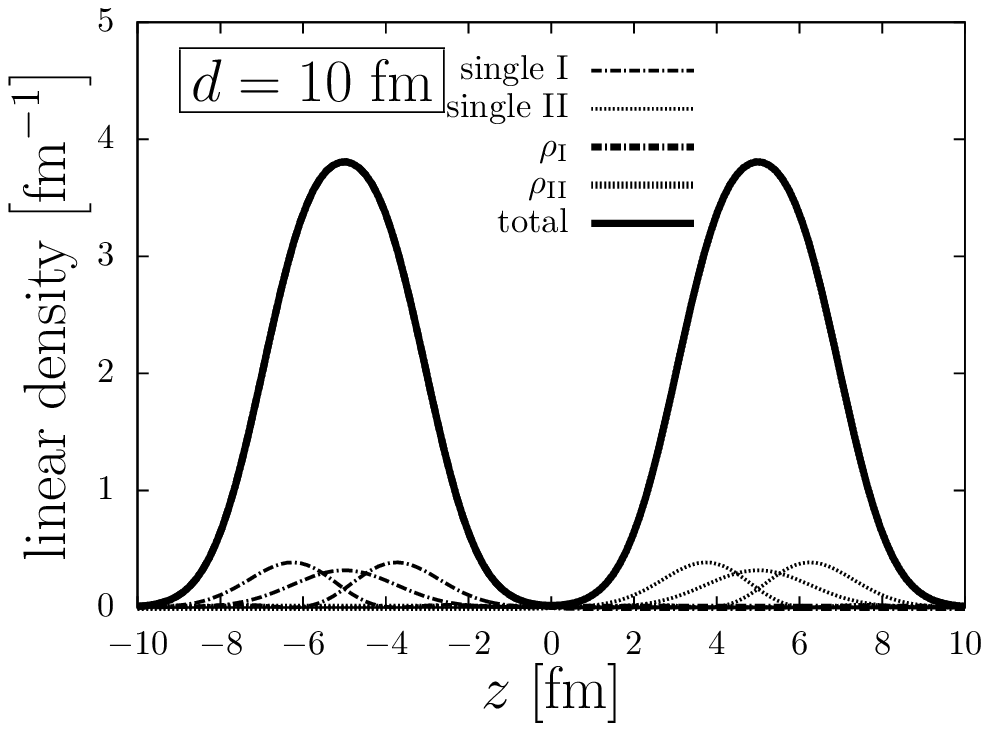}
       &
       \includegraphics[width=0.45\textwidth]{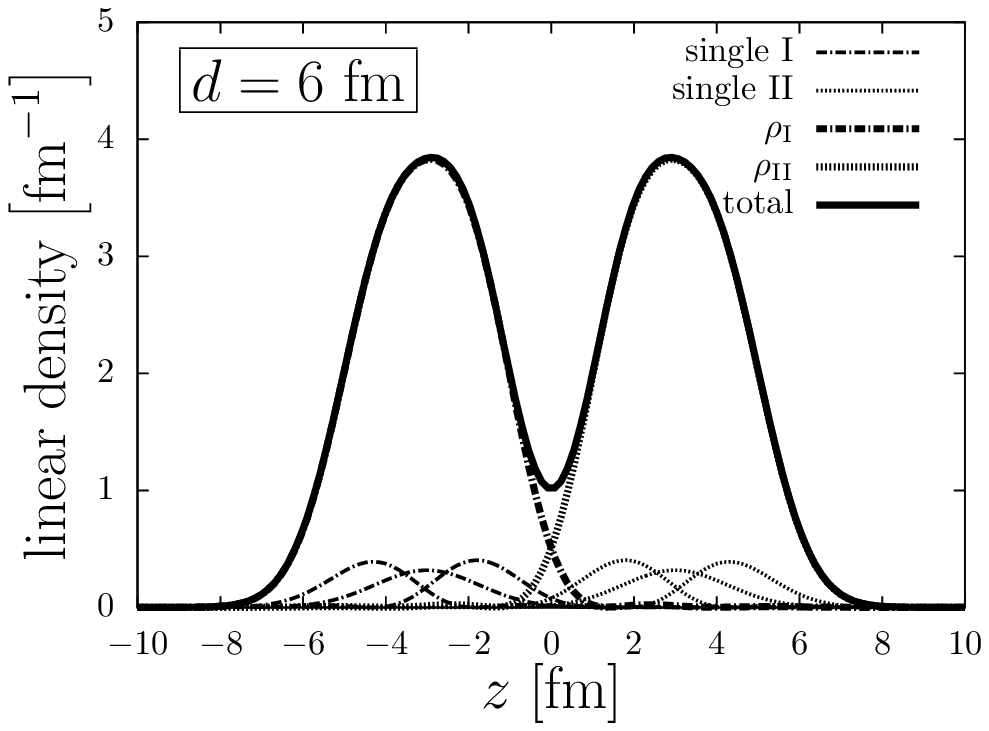}
       \\
       \includegraphics[width=0.45\textwidth]{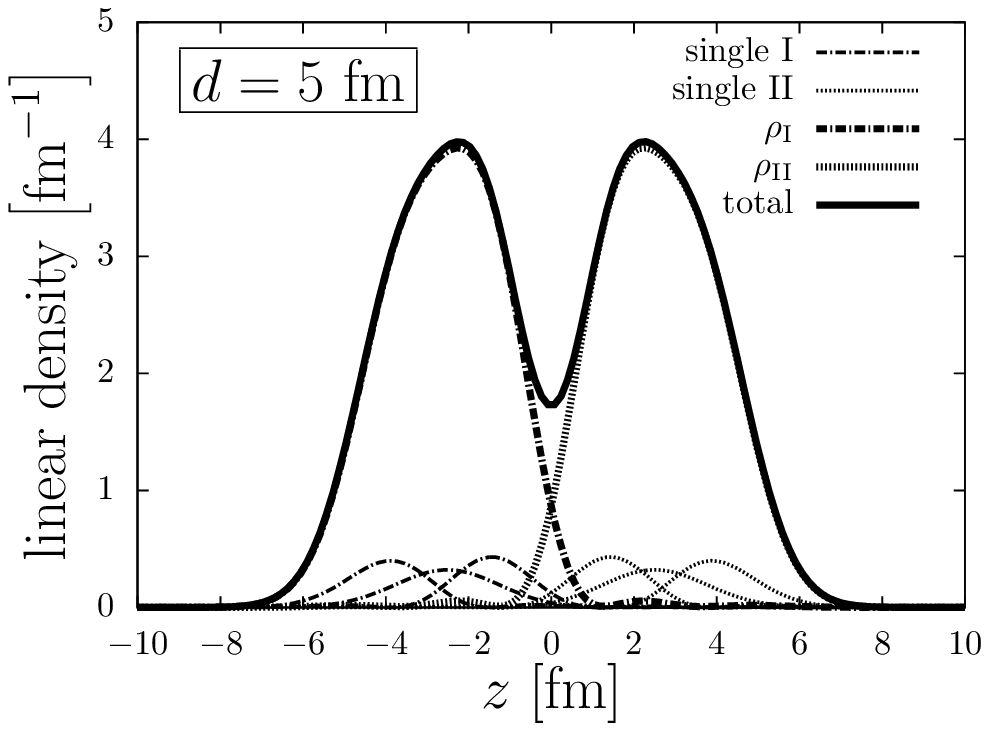}
       &
	   \includegraphics[width=0.45\textwidth]{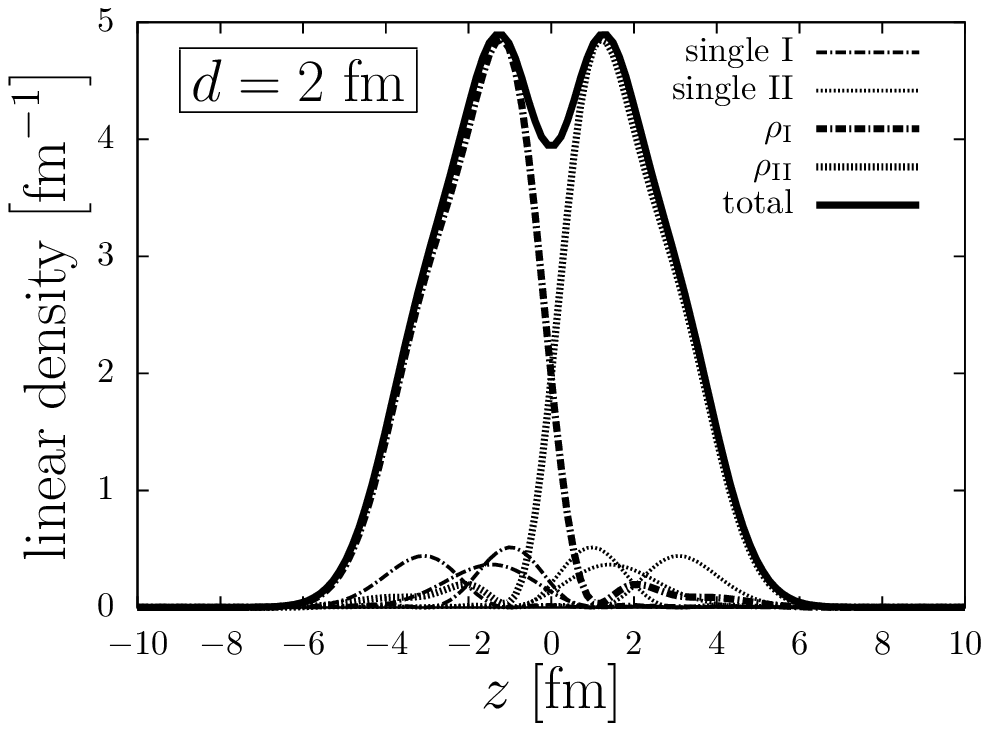}
	   \\
      \end{tabular}
      \caption{
      Linear density distributions of $^{16}$O clusters I and II, and the total system in $^{16}\mathrm{O} + ^{16}$O MB wave functions with the parameter $d= 10$, 6, 5, and 2 fm are plotted as functions of $z$.
      Thick dot-dashed and dotted lines represent single-particle wave functions for $^{16}$O clusters I and II, respectively.
      Thin dot-dashed and dotted lines represent $^{16}$O clusters I and II, respectively.
      Solid lines represent the total wave functions.
      Units of linear densities and $z$ are in fm$^{-1}$ and fm.
      }
      \label{density_OO}
     \end{center} 
    \end{figure}

Figure~\ref{density_OO} shows linear density distributions of $^{16}$O clusters and spatially localized single-particle wave functions in $^{16}$O + $^{16}$O MB wave functions with $d = 10,$ 6, 5, and 2 fm.
Width parameters $\nu$ are set to 0.16 fm$^{-2}$.
Each total density distribution shows a neck between two clusters, and each single-particle wave function is well localized.
Therefore, nucleons are separated into two $^{16}$O clusters.
The densities of the total system and the two $^{16}$O clusters are also shown in the figure.
In the touching region such as $d=5$ and 6 fm, 
the density profile for each $^{16}$O cluster defined in the present method 
also shows reasonable shapes. 
Subsystem densities have no singularity, and are smoothly damped in the overlap region. 

\begin{figure}[tbp]
 \begin{center} 
  \begin{tabular}{cc}
   \begin{minipage}{0.45\textwidth}
    \includegraphics[width=\textwidth]{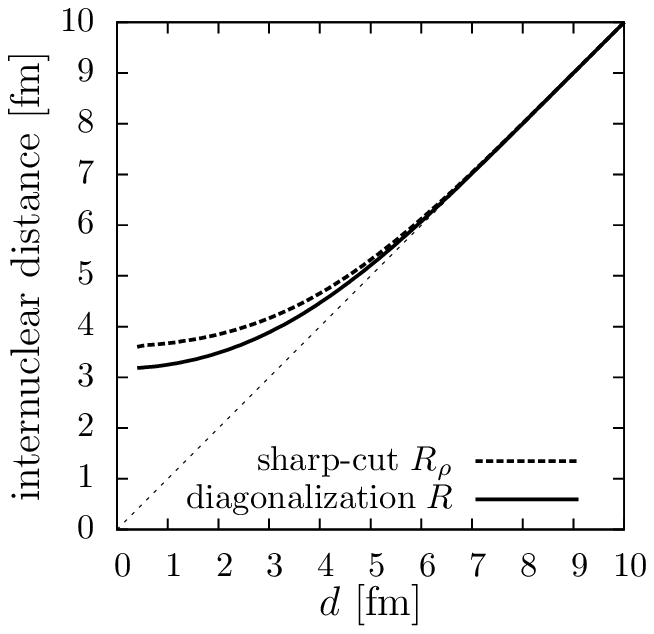}
    \caption{
    Internuclear distance between $^{16}$O clusters defined by the present method (solid) and sharp-cut separation (dashed) as functions of the distance between centroids of wave packets.
    Units are in fm.
    }
    \label{distance_OO}
   \end{minipage}&
   \begin{minipage}{0.45\textwidth}
    \includegraphics[width=\textwidth]{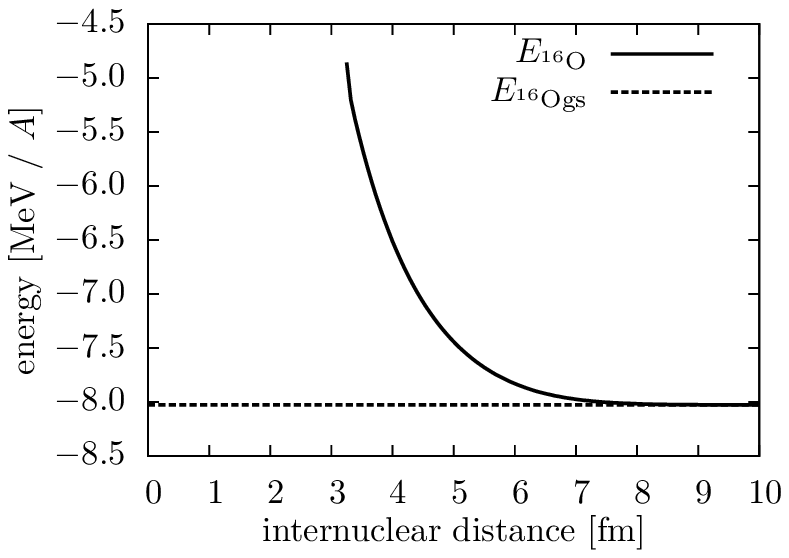}
    \caption{
    Energies of the $^{16}$O cluster in $^{32}$S are plotted as a function of internuclear distance in the unit of MeV/$A$ (solid), where $A$ is the mass number of the cluster.
    The dashed line shows the ground-state energy of the $^{16}$O cluster.
    }
    \label{energy_OO}
   \end{minipage}
  \end{tabular}     
 \end{center}
\end{figure} 

The internuclear distance $R$ determined by the present method is plotted as 
a function of $d$ in Fig.~\ref{distance_OO}. The distance $R_\rho$ defined by the sharp-cut density
is also shown for comparison.
$R$ equals to $d$ in the large $d$ region, and as $d$ decreases, 
it deviates from $d$ toward the $R>d$ region.
In the $d=0$ limit also, the internuclear distance $R$ has a finite value because of the Pauli blocking effect. The behavior of $R$ is qualitatively similar to that of 
$R_\rho$. However, quantitatively, 
$R$ is slightly smaller than $R_\rho$ because the left and right nuclei have 
an overlap density in the present definition instead of the sharp-cut density.

Figure~\ref{energy_OO} shows energies of the $^{16}$O clusters in the $^{32}$S as a function of internuclear distance $R$.
In $R \gtrsim 7$ fm region, the $^{16}$O clusters are in the ground state, but in $R \lesssim 7$ fm region, the $^{16}$O clusters are distorted by the antisymmetrization effect between clusters, and their excitation energies go up.

\subsection{$N \neq Z$: $^{10}$Be ($\alpha + ^6$He)}

In this section, the present method is applied to AMD wave functions for $N \neq Z$ nucleus $^{10}$Be that is calculated by energy variations after parity and angular momentum projection.\cite{10Be} 
In the AMD model, an intrinsic wave function is expressed by a Slater determinant of 
single-particle wave functions each of which is written by a Gaussian wave packet. 
Gaussian centers of all nucleons are treated as independently varying parameters, 
and therefore, no clusters are assumed. One of the advantages of the AMD
method is that the formation and/or breaking of clusters can be described in the
framework.
Since clusters are not assumed in the AMD model, a grouping of nucleons into two clusters is not necessarily obvious, unlike the MB cluster model wave functions. 

\begin{figure}[tbp]
\begin{center}
  \begin{tabular}{cc}
  \includegraphics[width=.45\textwidth]{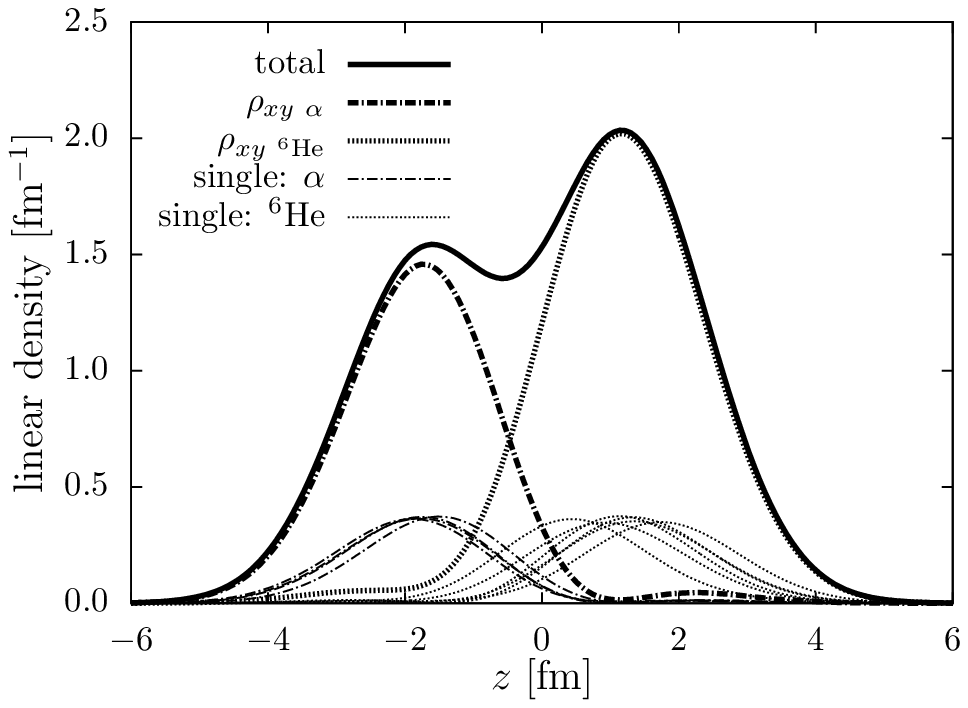}
  &
  \includegraphics[width=.45\textwidth]{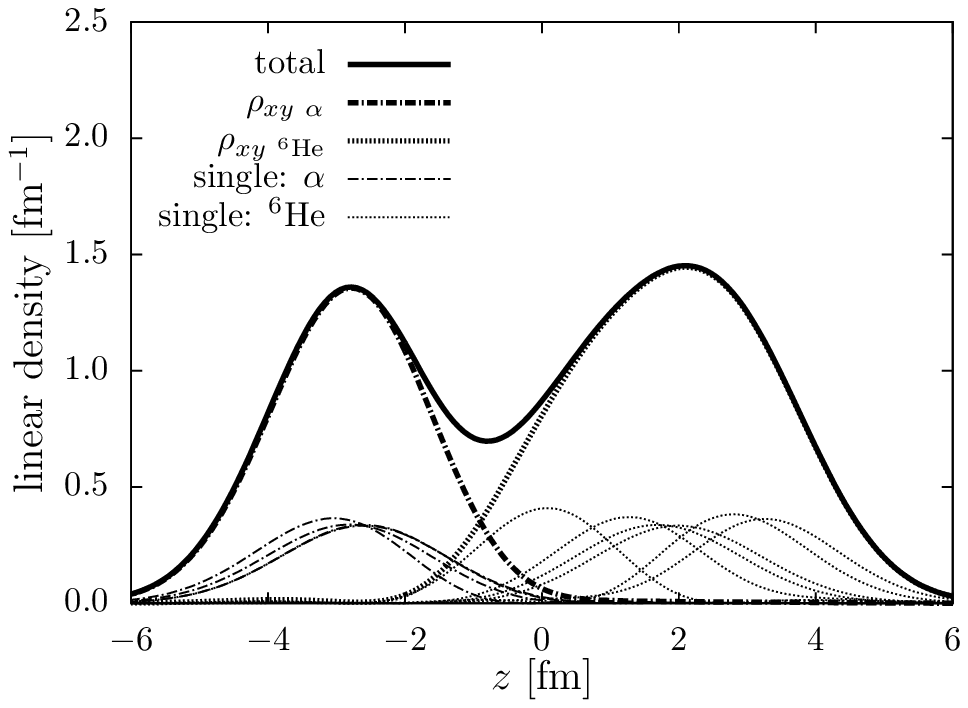}
  \\
 \end{tabular}
 \caption{
 Linear density distributions of intrinsic wave functions of dominant components of the $J^\pi = 0_1^+$ (left) and $0_2^+$ (right) states in $^{10}$Be.
 Thin dot-dashed and dotted lines represent single-particle wave functions for $\alpha$ and $^6$He clusters, respectively.
 Thick dot-dashed and dotted lines represent $\alpha$ and $^6$He clusters, respectively.
 Solid lines represent total wave functions.
 Units of linear densities and $z$ are in fm$^{-1}$ and fm.
 }
 \label{density_10Be}
\end{center}
\end{figure}

Analyzing the mean position $z_i$ of nucleons for intrinsic wave functions determined by
the AMD method, single-particle wave functions $\varphi_i$ can
be separated into two subsystems I and II. Four nucleons have $z_i$ 
in the left of the neck position and form an $\alpha$ cluster (subsystem I). Six nucleons
having $z_i$ in the right of the neck can be classified in subsystem II corresponding to 
a $^6$He cluster.
Figure~\ref{density_10Be} shows linear density distributions of intrinsic states of dominant components of  the $J^\pi = 0_1^+$ and $0_2^+$ states in $^{10}$Be, which are separated into $\alpha$ and $^{6}$He clusters.
In both states, total linear density distributions have necks at $z \sim 0$ fm, and single-particle orbits localize left and right parts. Therefore, the entire $^{10}$Be system can 
be separated into two clusters using the present separation method.

\section{Discussions}
\label{discussions}

The previous section shows that the present separation method involving the
diagonalization of a spatial operator of the major axis works well for the separation of  a Slater determinant wave function into spatially localized wave functions of subsystems.
The results of applications to MB wave functions with 
$\alpha$ + $\alpha$, $\alpha$ + $^{16}$O, and $^{16}$O + $^{16}$O structures are 
shown in Figs.~\ref{density_alpha_alpha}, \ref{density_AO}, and \ref{density_OO}, respectively.
With this method, we obtain subsystem wave functions with no singularity.
Because of non-singular wave functions, energies of subsystems do not diverge as shown in Figs.~\ref{energy_AA}, \ref{energy_AO}, and \ref{energy_OO}.
 This is one of the
advantages over the sharp-cut method.
Furthermore, the present method provides single-particle wave functions of subsystems, which are useful in the microscopic analysis of structures of subsystems,
and is applicable to studies of structural changes of subsystems 
in cluster structures and nuclear reactions, depending on the internuclear distance.
In the applications involving AMD wave functions of $J^\pi = 0_1^+$ and $0_2^+$ states in $^{10}$Be, 
although the existence of clusters is not assumed, 
the proposed method is proven to be useful in the separation of $^{10}$Be wave functions 
into subsystems $\alpha$ and $^6$He, when the neck positions and the 
single-particle orbits (Fig.~\ref{density_10Be}) are compared.
The proposed method is applicable to the separation of Slater determinant wave functions with neck structures into spatially localized subsystems.

As mentioned above, the present method gives wave functions of subsystems with which 
expectation values of operators can be calculated.
For example, the calculation of mass centers of subsystems determines the internuclear distances of $\alpha$ + $\alpha$, $\alpha$ + $^{16}$O, and $^{16}$O + $^{16}$O systems, 
as shown in Figs.~\ref{distance_2alpha}, \ref{distance_AO}, and \ref{distance_OO}, respectively.
The present method is useful to study cluster structures and nuclear reactions 
because the internuclear distance is an important degree of freedom in these phenomena.

    \begin{figure}[tbp]
    \begin{center}
     \includegraphics[width=0.5\textwidth]{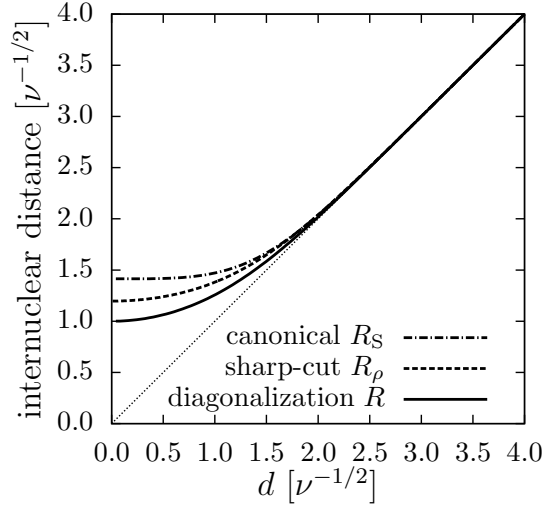}
    \end{center}
    \caption{Internuclear distance for the $\alpha + \alpha$ MB wave functions
specified by the parameter $d$.
The distance $R$ defined in the present method, $R_S$ defined in the S method, and $R_\rho$ defined by the sharp-cut density 
are plotted.
     Units are in $\nu^{-1/2}$.
   }
    \label{saracena}
  \end{figure} 

As an internuclear distance including the antisymmetrization effects between clusters, Saraceno \textit{et al.} proposed the internuclear distance as one of canonical variables for an $\alpha + \alpha$ system,\cite{saraceno}
 \begin{equation}
R_\mathrm{S} = \sqrt{\frac{1 + e^{- \nu d^2}}{1 - e^{- \nu d^2}}} d = \sqrt{1 + e^{- \nu d^2}} R.
 \end{equation}
We compare $R_\mathrm{S}$ with $R$ and discuss properties of these internuclear distance that include the antisymmetrization effects between clusters.
Although definitions of internuclear distances $R$ and $R_\mathrm{S}$ are different, the form of $R$ in the present definition resembles that of $R_\mathrm{S}$ in the denominator, which is related to the antisymmetrization effect,
but is  
smaller than $R_S$ because of the factor $\sqrt{1 + e^{- \nu d^2}}$ (Fig.~\ref{saracena}), and in the $d=0$ limit, $R=R_\mathrm{S}/\sqrt{2}$.
In the present method, single-particle orbits are defined to be completely orthogonal.
Under this condition, the internuclear distance is smaller than $R_\mathrm{S}$. 
With the increase in $d$, both the distances $R$ and $R_\mathrm{S}$ approach $d$. 

Advantages of the present method are as follows. 
First, subsystem wave functions can be obtained. They are expressed by an orthogonal set of single-particle orbits so that the Fermion feature of constituent nucleons is respected. The internuclear distance given by our method is the largest value under this condition. With the wave functions, subsystem properties such as the subsystem energy can be discussed. This is an advantage over the S method and the sharp-cut density method. 
Second, our method is applicable to any systems even in cases of proton- or neutron-excess systems or systems having one or more open-shell clusters or both. 
Third, our proposed method obtains spatially localized single-particle wave functions with no ambiguity in a simple linear transformation, which can be an advantage over the YG method.
In the YG method, the neck position has to be determined to obtain localized single-particle wave functions. The determination of the neck position may have 
less ambiguity in a case of a well separated system at a scission point, but 
it may have ambiguities in particular when two nuclei overlap to each other. Moreover, 
in a case of much proton- or neutron-rich systems, the neck position is unclear and neck positions of protons and neutrons can be different.

\section{Conclusions}
\label{conclusions}

In this paper, we proposed a method to separate a Slater determinant wave function with a neck structure into spatially localized subsystems.
Our method is applied to the MB wave functions of $\alpha$ + $\alpha$, $\alpha$ + $^{16}$O, and $^{16}$O + $^{16}$O systems, and the AMD wave functions of dominant components of the $J^\pi = 0_1^+$ and $0_2^+$ states in $^{10}$Be.

The proposed method obtains wave functions of subsystems with no singularity, which is an advantage over the sharp-cut method.
Using the obtained subsystem wave functions, we calculated the expectation values of several operators for 
subsystems. 
For example, the internuclear distance is well 
defined by the calculated mass centers of the subsystems, and energies of subsystems do not diverge.

The proposed method is simple and applicable to a wide variety of approaches that are
based on the Slater determinant wave function, e.g., the HF method.
It is useful for the analysis of systems that have spatially localized subsystems
in phenomena such as cluster structures and nuclear reactions.

\section*{Acknowledgments}

Numerical calculations were conducted on the high-performance computing system in the Research Center for Nuclear Physics, Osaka University. 
The authors thank Prof.~H.~Horiuchi and Dr.~M.~Kimura for fruitful discussions and valuable comments. 
This work was supported by the KAKENHI(C) 22540275.



\appendix
%
\section{}
The kinetic term $\hat{T}' \equiv \hat{T} - \hat{T}_\mathrm{G}$ with the center-of-mass correction is described as a two-body operator as follows,
\begin{eqnarray}
 \hat{T}' &\equiv& \hat{T} - \hat{T}_\mathrm{G} \nonumber\\
 &=& \sum_i \frac{\hat{\mathbf{p}}_i^2}{2m} - \frac{\left( \sum_i \hat{\mathbf{p}}_i\right)^2}{2Am} \nonumber \\
 &=& \frac{1}{2m} \sum_i \hat{\mathbf{p}}_i^2 - \frac{1}{2Am} \sum_{ij} \hat{\mathbf{p}}_i \cdot \hat{\mathbf{p}}_j \nonumber \\
 &=& \frac{1}{2Am} \sum_{ij} (\hat{\mathbf{p}}_i^2 - \hat{\mathbf{p}}_i \cdot \hat{\mathbf{p}}_j )\nonumber \\
 &=& \frac{1}{2Am} \sum_{ij} \left( \frac{\hat{\mathbf{p}}_i^2 + \hat{\mathbf{p}}_j^2}{2} - \hat{\mathbf{p}}_i \cdot \hat{\mathbf{p}}_j \right) \nonumber \\
 &=& \frac{1}{2Am} \sum_{ij} \frac{(\hat{\mathbf{p}}_i - \hat{\mathbf{p}}_j)^2}{2}  \nonumber \\
 &=& \sum_{i < j} \frac{(\hat{\mathbf{p}}_i - \hat{\mathbf{p}}_j)^2}{2Am}.
\end{eqnarray}

By a similar procedure, the mean squared radius $\bar{\mathbf{r}}^2$ with the center-of-mass correction is described as
\begin{equation}
 \bar{\mathbf{r}}^2 \equiv \frac{1}{A} \sum_i \left( \mathbf{r}_i - \frac{1}{A} \sum_j \mathbf{r}_j \right)^2 = \sum_{i < j} \left( \frac{\hat{\mathbf{r}}_i - \hat{\mathbf{r}}_j}{A} \right)^2.
\end{equation}

\end{document}